\documentclass[a4paper,10pt]{article}

\usepackage{latexsym}
\usepackage{amsfonts}
\usepackage{eucal}
\usepackage{amsmath}
\usepackage{latexsym}

\newenvironment{proof}[1][Proof]{\textbf{#1.} }{\ \rule{0.5em}{0.5em}}

\title{SHOR'S ALGORITHM FROM THE MINDSET OF QUANTUM ORACLES}
\author{L. Cimmino (\texttt{cimmino@na.infn.it})}
\begin{document}
\maketitle \vspace{0.5cm}
\begin{abstract}
The aim of this work is to show a brand-new way of making deterministic Quantum Computing (short QC), in the sense of Theory of Calculability, by meaning of unitary evolution.
We start from the original Shor's Algorithm to explain how the newest one works, at least compared to theory. We will give a new conceptual foundation of QC, resulting from a set of conventional and well known results of Calcolability and Quantum Mechanics. In the practice, if that can be used in its general sense, we will show an inaccessible relativized process which let us able to obtain same results with the same outlay in the time resource as the Shor's one for factorizing a given number $n$. Then the \textsl{Quantum Oracle} will be a prototype way giving to the relativized calculus the possibility to put in to practice an oracle, kind of object having till now abstract nature. 

The basic physical tool of our theorization, we call Quantum State Selection, consists in the twin-combined measurement process through \textsl{positive valued measure operator} (POVM)[Per02], needed to provide the quantum oracle's answer.
\end{abstract}

\section{Basic Ideas and Necessary Common Tools}
We want calculate the factors of a given number $n$. This kind of task is hard to solve cause the great amount of time resource needed to do it. Growing up the number $n$ will produce an exponential increase in the time resource of the process, depending by the tail of the input data. As one see, since Shor's algorithm was published, the quantum nature of the processes underling implies an exceptional decrease of time to polynomial magnitude.

One wants to find factors of the given number $n\in\mathbb{N}$. The Algorithm to do this maybe divided in three parts:
\begin{itemize}
	\item \texttt{first} - calculate the \textit{great common divisor} $(a,n)$ between $n$ and a random natural number $a<n$, the running time we consider is $O(n^2)$. I've newly noticed there are complex conventional algorithms taking time of $O(n(\log n)^2\log\log n)$
	\item \texttt{second} - if $a$ is coprime to $n$, then find the period $r$ of the \textit{reminder function} $f(x)=a^x\mod n$, so such that $f(x+r)=f(x)$, $\forall x\in \mathbb{N}$, where the period or order is the least integer $r$ such that $y^r\equiv 1 \mod n$
	\item \texttt{finally} - if period was found and it is even, then calculate $(a^{r/2}\pm 1,n)$ giving at least one factor of $n$
\end{itemize}
The total amount of time resource is of order $O(n^2) + x + O(n^2)$, where the order of the third part is been evalueted as of order $O(n^2)$, since it is calculated by running Euclid's algorithm one time only. Now, $x$ represents the order due to the run of the \textit{finding period} sub and since we know we are working conventionally with an hard task, we expect to find a process allowing us to compute efficiently this part of the Shor's algorithm. So conventionally we have $x=O(n^n)$ and quantistically we want to find $x = O(n^p)$, were $p$ is any non-negative integer.

\vspace{.5cm}

In a modern non-original form, the Euclidean algorithm acts same as 
\vspace{.2cm}
\newline\texttt{1\hspace{0.3cm}	int gcd(int a, int b)} $\{$
\newline\texttt{2\hspace{1.3cm}		return ( b != 0 ? gcd(b , a \% b) : a );}
\newline\texttt{3\hspace{0.3cm}	}$\}$
\vspace{.2cm}
\newline where \texttt{a \% b} is the reminder of the division $\frac{a}{b}$ and the procedure return the value found in the first register \texttt{a} if the second argument of the function \texttt{cgd} equals 0, recurse otherwise. For example, by putting \texttt{a=110} and \texttt{b=129}, one has
\vspace{.1cm}
\newline\texttt{\#\hspace{0.5cm}	a \hspace{0.5cm} b}
\newline\texttt{1\hspace{0.3cm}	110\hspace{0.3cm} 129}
\newline\texttt{2\hspace{0.3cm}	129\hspace{0.3cm} 110}
\newline\texttt{3\hspace{0.3cm}	110\hspace{0.5cm} 19}
\newline\texttt{4\hspace{0.5cm}	19\hspace{0.5cm} 15}
\newline\texttt{5\hspace{0.5cm}	15\hspace{0.7cm} 4}
\newline\texttt{6\hspace{0.7cm}	4\hspace{0.7cm} 3}
\newline\texttt{7\hspace{0.7cm}	3\hspace{0.7cm} 1}
\newline\texttt{8\hspace{0.7cm}	1\hspace{0.7cm} 0}
\vspace{.2cm}
\newline
and since \texttt{a = 1} it results that $110$ and $129$ are coprime.

For what follows, it is useful notice that by repeating $m$ times the Euclidean algorithm, once $n$ is fixed and keeping $m<n$ for all $n$, one has a process runs in $m\cdot O(n^2)$, hence increasing the algorithm's time resource by a multiplicative factor polynomial in $n$, or by assuming the count of $m$ of order $O(n)$, we fix the upper bound to the order $O(n^3)$.

\vspace{.5cm}
Now, we show how $(a^{r/2}\pm 1,n)$ gives us at least one factor of $n$ by considering we found, in the second stage of the algorithm, the period $r$ being an even number. Let $a, n\in\mathbb{N}$ such that $a<n$ and $(a,n)=1$, then $(a^{r/2}+ 1,n)$ or $(a^{r/2}- 1,n)$ divides $n$. 
\begin{proof}
Taking into account the \textit{reminder function} $f(x)=\{\frac{a^x}{n}\}$ one has $\{\frac{a^{x+r}}{n}\}=\{\frac{a^x}{n}\}=\eta$. So 
\begin{eqnarray}
a^xa^r&\equiv\eta\mod n\nonumber\\
a^x&\equiv\eta\mod n\nonumber
\end{eqnarray}
then subtracting member by member the two modular congruencies, it results
\begin{eqnarray}
a^x(a^r-1)\equiv 0\mod n\nonumber
\end{eqnarray}
and finally one has $a^r-1\equiv 0\mod n$ or $\{\frac{a^r-1}{n}\}=0$.
Since $r$ is odd and $a^r-1=(a^\frac{r}{2}+1)(a^\frac{r}{2}-1)$, one finds
\begin{eqnarray}
\Big\{\frac{(a^\frac{r}{2}+1)(a^\frac{r}{2}-1)}{n}\Big\}=0\iff\Big\{\frac{a^\frac{r}{2}+1}{n}\Big\}=0\hspace{.2cm}\mbox{or}\hspace{.2cm}\Big\{\frac{a^\frac{r}{2}-1}{n}\Big\}=0.\nonumber
\end{eqnarray}
\end{proof}

The latter runs of order $O(n^2)$ in the time resource and it maybe executed at most twice per time. Stated what we shown till now if the second stage runs in polynomial time order we will have a process that runs in the same magnitude in time resource. This question has already been answered positively as testified in the masterpiece of paper about QC. In the next section we will explain briefly how Shor's algorithm acts and how can be evaluated the order of time complessity required by it.

\section{Quantum Part of Shor's Algorithm}
By Euclidean algorithm we found an $a$ coprime to $n$, so one can compute the order or the reminder function $a \mod n$. Tacking into account the Jozsa picture [Joz98] and start with the state $|\hspace{0.1cm}0\hspace{0.1cm}\rangle|\hspace{0.1cm}0\hspace{0.1cm}\rangle$, since we are working with a two-register system, a simplified scheme for quantum procedure consists of the following steps:
\begin{enumerate}
	\item Apply the unitary operator $A_q$ such that $A_q|\hspace{0.1cm}0\hspace{0.1cm}\rangle|\hspace{0.1cm}0\hspace{0.1cm}\rangle=\sum_{x=0}^q{|\hspace{0.1cm}x\hspace{0.1cm}\rangle}|\hspace{0.1cm}0\hspace{0.1cm}\rangle$, where $q=2^l$, being $l$ the number of digits in use.
	\item Apply the $U_f$ unitary transformation to the previous state second register to obtain the state $\sum{|\hspace{0.1cm}x\hspace{0.1cm}\rangle|\hspace{0.1cm}a^x\mod n\hspace{0.1cm}\rangle}$. This task is obscure cause it seems evocate some kind of quantum alchemy by meaning of an 'oracle problem'; Shor [Sho96] refers to it as a kind of black box subroutine, whose code is inaccessible. Later, we will back on this argument and we'll discus the solution.
	\item Apply another time $A_q$ operator on the first register obtaining the state $\sum{e^{2\pi i\frac{tx}{q}}|\hspace{0.1cm}t\hspace{0.1cm}\rangle|\hspace{0.1cm}a^x\mod n\hspace{0.1cm}\rangle}$, where it results $t=\frac{xq}{r}$.	
\end{enumerate}
Now before step 3 happens, one measures the second register to keep the corresponding state of the first one, and after it, one performs a measurement on the first register. 

Since $t$ is measured and $q$ is fixed, by performing the procedure, including the sequence of measurements, as much as it is enough to determine a $t$ coprime to $q$, one has by continued fraction the value of $r$ by means of canceling the fraction at the second term of $\frac{x}{r}=\frac{t}{q}$ to the lowest terms.

A collaboration between IBM and Stanford University's researchers [Chu01] has shown how this can be made using a spin system. To factorize $n=15$ they build a system of seven qubit molecules, based upon the consideration that one needs four qubit (second register) to enter data about $a^x\mod n$ and three qubit (first register) to evaluate the period.  Fixing $a=11$ and by interpretation of experimental data they found the first register in a equal mixture of states $|\hspace{0.1cm}0\hspace{0.1cm}0\hspace{0.1cm}0\hspace{0.1cm}\rangle\equiv|\hspace{0.1cm}0\hspace{0.1cm}\rangle$ and $|\hspace{0.1cm}1\hspace{0.1cm}0\hspace{0.1cm}0\hspace{0.1cm}\rangle\equiv|\hspace{0.1cm}4\hspace{0.1cm}\rangle$, so that the period is found by the relation $r=2^3/4=2$ thus obtaining $(11^{2/2}\pm 1,15)=3_{(+)}$ and $5_{(-)}$. Similarly, they fixed $a=7$, getting an equal mixture of states $|\hspace{0.1cm}0\hspace{0.1cm}0\hspace{0.1cm}0\hspace{0.1cm}\rangle\equiv|\hspace{0.1cm}0\hspace{0.1cm}\rangle$, $|\hspace{0.1cm}0\hspace{0.1cm}1\hspace{0.1cm}0\hspace{0.1cm}\rangle\equiv|\hspace{0.1cm}2\hspace{0.1cm}\rangle$, $|\hspace{0.1cm}1\hspace{0.1cm}0\hspace{0.1cm}0\hspace{0.1cm}\rangle\equiv|\hspace{0.1cm}4\hspace{0.1cm}\rangle$ and $|\hspace{0.1cm}1\hspace{0.1cm}1\hspace{0.1cm}0\hspace{0.1cm}\rangle\equiv|\hspace{0.1cm}6\hspace{0.1cm}\rangle$, thus it resulted $r=4$ so that $(7^{4/2}\pm 1,15)=$ $=3_{(-)}$ and $5_{(+)}$. Further, as it has been shown by Jotza [Joz97] and Hoyer [Hoy97], Shor's algorithm runs in polynomial time.

\section{Period Finding in the \textsl{Quantum Oracle} System picture}
As shown, Shor's algorithm constitutes the right way to run fast factorizing a given number $n$. But as predate before, in literature the use of oracle is obscure and, some times, rough. From theory of calcolability, an oracle is associated to a set of integer numbers, whose characteristic function is self-calculated by mean of relativized calculus. In other words can not exists a cognizable formal process evaluating all elements of the oracle, its task is just to answer about the presence in it of a given number or not, through interrogation by the side of Turing Machine. Unlike QC till now, we want a conventional relativized calculus that joins the needs to put in practice an object such that.

\subsection*{\textit{Notation and critics to oracle subroutine}}
First, one acts producing a finite superposition in the first register by means of Shor's algorithm step 1. Using Barenco notation [Bar95], the operator $A_q$ can be expressed by the matrix
\begin{multline}
A_q\equiv H_{l-1}\bigwedge_1(U_{l-2,l-3})H_{l-2}\bigwedge_1(U_{l-3,l-1})\bigwedge_1(U_{l-3,l-2})\\
H_{l-3}\bigwedge_1(U_{l-4,l-1})\bigwedge_1(U_{l-4,l-2})\bigwedge_1(U_{l-4,l-3})R_{l-4}\ldots\\
\ldots H_1\bigwedge_1(U_{0,l-1})\bigwedge_1(U_{0,l-2})\ldots
\bigwedge_1(U_{0,1})\bigwedge_1(U_{0,0})H_0\nonumber
\end{multline}
being
\[
H_j=\frac{1}{\sqrt 2}
\begin{pmatrix}
    1 & 1\\
    1 & -1
\end{pmatrix}\hspace{1cm}
U_{jk}=
\begin{pmatrix}
    1 & 0\\
    0 & e^{i\theta_{k-j}}
\end{pmatrix}
\]
referring the indexes to the corresponding qubit in the register.
So, starting from initial first register $|\hspace{0.1cm}0\hspace{0.1cm}\rangle$ setting, one reaches the state $\sum_x{|\hspace{0.1cm}x\hspace{0.1cm}\rangle}$. The second register is quite untouched by this operation and now it will be left to evolve under the 'dedicated-like' transformation $U_f$.

The task of modular exponentation, involving $U_f$, is conventionally the hardest one in the Shor's picture and it constitutes the bottle-neck of the process [Sho9]. As it was anticipated, this is the obscure part of the entire process; Shor speaks about reversible calculus to do quantum modular exponentation, having in mind the fact settling '\textit{Because of the reversibility of quantum computation, a deterministic computation is performable on a quantum computer only if it is reversible.}' and it progresses through a quantum version of the modular exponentation algorithm.
To criticize this step is vain, cause the absurd idea of making QC on a quantum computer it isn't actually possible; on the contrary, what's to emphasize, it is the absolute misunderstanding about what concerns to oracle and what concerns to the quantum machine. Since an Oracle is an arbitrary entity, associated to a set of natural number which can answer yes or no to a belonging-question, i.e. '\textit{does the given number belongs to the set \textsl{Q}?}'. But, if someone asks what's an oracle, we just answer that it is not a black-box, not a subroutine, not an formal process letting us able to know it. So the quantum part of the process appears only in the moment of interrogation by the side of the Turing Machine. Neither measure, nor probability, nor indeterminate results, the aim of this kind of quantum system is to answer the question letting the TM able to proceed its evolution. But while we can think about a quantum computer and its logical gates, as an expansion of conventional digital electronics based computer, more fast, more efficient, dissipating a minor amount of heat or at most heat free, the nature of this entity must be compared to quantum mechanics. The oracle's nature is quite different, just like its semantic is apart from the meaning of formal computation, nevertheless it can embrace the idea of determinism. Keeping away ourselves from field of technology which someone could claim to be \textit{quantronics}, we state that what we could be able to employ in the quantum computers and what we want to do implementing the oracle entity, substantially differ, in the final stage, for the kind of measure used at the end of the process. In the following we will shape a Quantum Machine Model, we think realizable.

\subsection*{\textit{Description}}
A \textsl{QO}-System is a relativized Turing Machine whose oracle $Q$ is a quantum answering system. Beyond Shor's picture, the quantum part of the \textsl{QO}-System consists in making a start state, measuring it to select a corrensponding mixture of the second register depending by relation between the answer and the first register, for what's conventionally argued by quantum theory, evaluating the quantum data in second register and, finally, measuring by POVM the final state. What one obtains, it is a inaccessible effective process, where the inaccessible square with not knowable in the sense of quantum theory and effective with determined by initial data in the sense of theory of calcolability. 

Intuitively, the \textsl{QO}-System for factorization is a Turing Machine that acts in the following way:
\begin{enumerate}
	\item Given the $n$ to factorize, calculate through Euclidean algorithm the array of number $\{a_i\in\mathbb{N}:(a_i,n)=1, a_i<n\}$. This step runs of order $O(n^3)$ at most.
	\item Construct the initial state, in the same way it occurs in the Shor's picture. $|\hspace{0.1cm}Q\hspace{0.1cm}\rangle\equiv\sum_{r=1}^{n-1}{\bigotimes_{j=1}^h{e^{-i\omega r}|\hspace{0.1cm}r\hspace{0.1cm}\rangle|\hspace{0.1cm}a_j^r\mod n\hspace{0.1cm}\rangle}}$ is obtained acting with
	unitary operator $A_q$ on the first register such that $A_q|\hspace{0.1cm}0\hspace{0.1cm}\rangle|\hspace{0.1cm}0\hspace{0.1cm}\rangle=\sum{|\hspace{0.1cm}r\hspace{0.1cm}\rangle}|\hspace{0.1cm}0\hspace{0.1cm}\rangle$ and with unitary transformation $U_f$ on the second $h$-register to obtain the state $\sum_{r}\bigotimes_{j}{|\hspace{0.1cm}r\hspace{0.1cm}\rangle|\hspace{0.1cm}a_j^r\mod n\hspace{0.1cm}\rangle}$.
	\item The state thus obtained is a superposition of number state and it is \emph{telePOVM} from the internal entity \textsc{a} to the external \textsc{b} which selecting the state by use of information now stored on the Turing machine's tape, will force, by meaning of selection of quantum state, the first register to be $|\hspace{0.1cm}r_0\hspace{0.1cm}\rangle$.
	\item The question '\textit{Does $r_0$ belong to $Q$?}' will be answered, setting the $h$ registers storing the $f(x)$ into tensor product $\bigotimes_{j=1}^h|\hspace{0.1cm}a_j^{r_0}\mod n\hspace{0.1cm}\rangle$ caused by entangled relation with the first register now storing $|\hspace{0.1cm}r_0\hspace{0.1cm}\rangle$.   
\end{enumerate}

Since, at the end of step 4, the information is collapsed to the value of the modular function, we can read the exiting result by use of a quantum circuit, or maybe digital if one wants, made of NOT and XOR ports as it is illustrated in figure. So, if the output is set to 0 then the answer is affirmative and the TM is able to evaluate the factor of $n$, else the process restart at step 2 putting another $r_0$ into question to $Q$. 

\subsection*{\textit{Physical draft}}
We already claimed that in our model step 2 produces a superposition more complex than the Shor's one. This kind superposition can be studied through the theory of multientanglement [Lin98], but, without lose of generality, we intend to simplify our explanation referring it to the this simple case. 

The Step 3 lies in the mathematics of teleportation process for what's useful to our purposes. We refer to [Hug93] for what concerns the theory of $\rho-$ensembles and telePOVM starting point. The telePOVM consists in the teleportation of a $\rho-$ensemble by operating a POVM in the sending location. This fact arises on the equivalence between POVM and $\rho-$ensembles expressed by [Hug93], which let the sender \textsc{a} able to to choose the $\rho-$ensemble that will be received by \textsc{b}. 

In the present model the telePOVM process is used to bring the \textsl{QO}-system to be in the state of our interest, selecting the required state in the \textsc{b} location and, without loss of generality, we'll refer to single qubit transformation. So, following the scheme in [Mor99], one has an EPR pair shared by \textsc{a} and \textsc{b} and since we are dealing with POVM the number of results is larger than the dimension of the Hilbert space by the side of both sharers.

Before we observe its registers, from step 2, the machine is found in the superposition
\begin{equation}
\sum_{r}\bigotimes_{j}{|\hspace{0.1cm}r\hspace{0.1cm}\rangle|\hspace{0.1cm}a_j^r\mod n\hspace{0.1cm}\rangle}\nonumber
\end{equation}
so considering qubit-by-qubit the primary register $|\hspace{0.1cm}r\hspace{0.1cm}\rangle$, we deal with a tetrapartite system composed by four qubit, one of those associated to the \textit{question} and represented by the $i-$th one of \textit{ancilla}, the other one consisting of a couple of entangled qubit, respectively \textsc{a} and \textsc{b}, and the $i$-th one of the first register $|\hspace{0.1cm}r\hspace{0.1cm}\rangle$.

The mechanic of quantum state selection is characterized by the use of two measure, executed on both side of the couple, once every member, respectively, had interacted by means of suitable unitary operators, \textsc{a} with the $i-$th qubit of the register to be selected and \textsc{b} with the $i-$th qubit of the ancilla, or the \textit{selector}. The measurements are sequential, first on the \textsc{a} side, then on the \textsc{b} side and consists, in order, in a Bell measurements and in a PVM. This produce the \textit{selected} number state $r_0$, we're asking for, in the first register, that sets the secondary register to the corresponding value of remainder function. 
  
\section{Conclusions}
The emerging calcolability model, based on the operation of the quantum state selection, gives us the possibility to put in practice a machine able to execute quantum algorithms, as the Shor's one. This progress could be crucial to go on over the algorithmic solution of Nature. On the other hand, the quantum state selection, whose details can be found in [Cim10], based on the laurea thesis ``\textit{Modello Quantomeccanico ad un Numero Finito di Livelli per il Calcolatore Quantistico}'', if will be shown experimentaly, it could constitute a feasible advance to the theory of quantum measurements, upholding the primary role of entanglement.

\end{document}